%% file: paper.tex
\documentclass[reprint,aps,prb,amsmath,amssymb,nofootinbib,longbibliography]{revtex4-2}

\usepackage{graphicx}
\usepackage{booktabs}
\usepackage{dcolumn}
\newcolumntype{d}{D{.}{.}{-1}}
\usepackage{siunitx}
\usepackage[version=4]{mhchem}
\usepackage{microtype}
\usepackage{xcolor}
\definecolor{linkblue}{RGB}{20,45,110}
\usepackage{hyperref}
\hypersetup{colorlinks=true,linkcolor=linkblue,citecolor=linkblue,urlcolor=linkblue}

\DeclareSIUnit{\rydberg}{Ry}
\DeclareSIUnit{\bohr}{\ensuremath{a_0}}

\newcommand{\vlit}{V_{\mathrm{lit}}}
\newcommand{\vpred}{V_{\mathrm{pred}}}
\newcommand{\vmp}{V_{\mathrm{MP}}}

\begin{document}

\title{Computational references are not experiments: pre-registered validation of machine-learned sodium-cathode voltages}

\author{Krishna Teja Vepa}
\email[]{tejakrishnna@gmail.com}
\affiliation{Independent Researcher}
\altaffiliation{ORCID: 0009-0002-2400-2281}

\date{\today}

\begin{abstract}
Machine-learning screens for battery materials are trained and judged almost
entirely against computed reference voltages, and those references carry their
own systematic errors. We report a case in which this matters quantitatively:
our own screening stack (a graph-network voltage screen, a prior-art triage
layer, and a local PBE+$U$ bench) fails pre-registered validation against
experiment-anchored literature values. Verdict thresholds, failure modes, and
the primary metric were committed before analysis. On an operator-audited set
of known Na-ion cathodes ($n=6$ after one documented exclusion; verdict
unchanged at $n=7$), the raw held-out mean absolute error was \SI{0.67}{\volt},
the pre-registered conservative metric, the upper 95\% confidence bound of
the cross-validated bias-corrected error, was \SI{1.09}{\volt}, and the
residual was strongly voltage-dependent ($r=-0.94$), so no additive calibration
is valid. On the two compounds where prediction, database reference, and
experiment could all be compared, the Materials Project PBE+$U$ reference sat
about \SI{0.54}{\volt} below measurement: the reference, not the model,
dominated the error. A prior-art screen found at least 70\% of the targeted Na
substitution space already published. We retire the screen, bound what
``verified'' means for our DFT ledger, and run a pre-registered calibration
audit of it against four benchmark Li couples. That audit returns a standard
deviation of \SI{0.31}{\volt}, above its \SI{0.30}{\volt} bar, so absolute-voltage
claims are retired from our own ledger as well; the residual is chemistry-structured
(the layered-\ce{Co} and \ce{Fe}-phosphate couples share $+\SI{0.41}{\volt}$ to
within \SI{0.09}{\volt}, with the \ce{Mn} spinel apart), pointing to a per-chemistry
offset rather than a single global one.
\end{abstract}

\maketitle

\section{Introduction}
\label{sec:intro}

Generative models now propose inorganic crystals far faster than any
laboratory, or any density-functional-theory (DFT) queue, can check them. The
GNoME release alone reported 2.2 million candidate structures and declared
roughly 380{,}000 of them stable \cite{merchant2023}, and an autonomous
laboratory reported synthesizing dozens of those candidates within weeks
\cite{szymanski2023}. The published record since then has been less kind.
Cheetham and Seshadri found that almost none of the examined GNoME compounds
were simultaneously credible, new, and useful by the standards a chemist would
apply \cite{cheetham2024}, and a re-examination of the autonomous-synthesis
claims concluded that none of them was convincingly demonstrated
\cite{leeman2024}. The field's binding constraint has moved from generating
candidates to deciding which claims about them deserve trust.

That decision rests on reference data more often than the literature
acknowledges. Screening models for battery cathodes are typically trained on
intercalation voltages computed within GGA or GGA+$U$
\cite{aydinol1997,wang2006}, most commonly the Materials Project values
\cite{jain2013}, and they are usually \emph{evaluated} against the same
computed quantities. A model can therefore pass every benchmark available to
it while inheriting, invisibly, whatever systematic error separates its
reference scale from the electrochemical measurements the field actually
cares about. Validation against experiment is the only test that can expose
this, and it is the test least often run.

This paper runs that test on our own system, under rules fixed in advance.
The system under test (Sec.~\ref{sec:system}) is a small closed-loop materials
pipeline (generator, graph-network voltage screen, prior-art triage layer,
and a local PBE+$U$ verification bench) built and operated by one person.
Before any validation metric was computed, we committed verdict thresholds, a
conservative primary metric, and four named failure modes to the repository,
and a fifth stop condition was fixed before literature curation began, so
that the result could not be renegotiated after the fact
(Sec.~\ref{sec:protocol}); the commit hashes are the audit trail
\cite{nosek2018}. The outcome is negative on every axis we pre-registered.
The voltage screen is not screening-grade against experiment, and its residual
structure forbids the additive recalibration that would otherwise be the
obvious patch (Sec.~\ref{sec:resultsA}). The computed references themselves
sit about half a volt below measurement on the rows where a three-way
comparison was possible, which redistributes blame in an uncomfortable
direction: part of what the model learned to reproduce was reference error,
not electrochemistry (Sec.~\ref{sec:resultsB}). And the targeted ``discovery''
space the generator was aimed at was at least 70\% already published, before
any question of model accuracy arises. The constructive output is the
protocol itself, a small quote-anchored experimental reference set with
documented provenance, and a pre-registered calibration audit of our own DFT
scale, whose decision rule was committed in advance and whose verdict is
reported in Sec.~\ref{sec:resultsC}.

We claim no discovered material, no validated screen, and no calibrated
scale. The paper's contribution is the discipline: every quantitative claim
below traces to a committed artifact, every failure is reported under the
definition fixed before the analysis ran, and the strongest objections we
know of are stated and answered in Sec.~\ref{sec:discussion}.

\section{System under test}
\label{sec:system}

The pipeline, called QME, is a solo-operated closed loop that runs on a single
Apple M4 desktop. One cycle proceeds: generate candidate crystals, screen them
with a property-predicting graph neural network (GNN), rank survivors with a
Gaussian-process acquisition function over learned embeddings, screen the
ranked list for prior art, verify the top picks with paired DFT calculations,
and fine-tune the GNN on the newly verified results. Each stage is described
here at the level needed to interpret the validation; computational details
are in Sec.~\ref{sec:methods}.

Candidates come from prototype substitution: redox-site substitution into
known framework structures retrieved from the Materials Project (MP)
\cite{jain2013}, followed by charge-neutrality and electronegativity screening
with SMACT \cite{davies2019}. The screen is a Siamese multi-head GNN that
ingests a (charged, discharged) structure pair and predicts the average
intercalation voltage with a Monte-Carlo-dropout uncertainty. Its training
corpus contains 2814 graphs, of which 20 pairs (0.7\%) are Na
chemistries; the model is, by construction, a Li-dominated predictor asked to
extrapolate. A second network, used for stability estimates and for the
embedding space, is not ours: it is the universal interatomic-potential
backbone MACE-MP-0 \cite{batatia2022,batatia2024} with property heads, and we
treat it throughout as borrowed machinery. The acquisition function is a
plain upper-confidence-bound rule over a PCA reduction of those embeddings; a
diversity-weighted variant exists in the codebase but was never wired into the
orchestrator, and with five verified training points the acquisition stage has
no demonstrated value of any kind. We make no claim for it.

The prior-art layer is the part of the stack built to say \emph{no}. A
candidate is flagged when a literature, structure-database, or family-pattern
source names its composition: citations count only if a chemical formula in
the retrieved record matches the candidate's composition exactly, as a
de-intercalated framework, or as the same element set, which removes the
loose-token false positives that plague relevance-ranked search. The layer is
safety-locked: it can downgrade a candidate to known but can never certify
novelty, and its machine verdict for absence is bounded by the coverage of the
sources it actually searched. In its committed precision audit it flagged
8 of 8 known compounds and false-flagged 0 of 5 fictional control
compositions.

Verification is paired-state DFT: both members of each (charged, discharged)
pair are relaxed under identical PBE+$U$ settings with Quantum ESPRESSO
\cite{giannozzi2009,giannozzi2017}, and the average voltage follows from the
total-energy difference (Sec.~\ref{sec:methods}). The verified ledger
contains seven rows: five known Li-ion cathodes that serve as calibration
anchors (\ce{LiCoO2} at \SI{4.120}{\volt}, \ce{LiFe(PO3)4} at
\SI{5.687}{\volt}, \ce{LiNiP2O7} at \SI{5.232}{\volt}, \ce{LiFeP2O7} at
\SI{5.275}{\volt}, \ce{Li2Fe(PO3)5} at \SI{5.6115}{\volt}), one experimental
touchpoint, and one computed-only couple with no experimentally anchorable
counterpart (Appendix~\ref{app:ledger}). Every row is voltage-only: hull
distances were not computed and kinetic screens were never run, so
``verified'' in this paper always means a converged voltage on a consistent
PBE+$U$ scale and nothing more. Four of the five anchors are
metaphosphate or pyrophosphate couples between \SI{5.2}{\volt} and
\SI{5.7}{\volt}, outside any practical electrolyte window and, to our
knowledge, without experimental electrochemistry at those couples. The
ledger's single experimental touchpoint is \ce{Na3V2(PO4)3}, computed at
\SI{2.8957}{\volt} against a measured plateau near \SI{3.4}{\volt}
\cite{jian2012}. Zero candidates classified as absent from our corpora
have ever been verified, so nothing in the ledger tests discovery.

The loop has closed once: the fifth anchor triggered a fine-tune of the GNN,
gated on a held-out leave-one-out cross-validation over the five anchors and
on the fitted anchor error (\SI{0.084}{\volt} mean absolute), and the
fine-tuned model was activated and used for every prediction in this paper.
Both gates compare the model to its own DFT anchors. Whether that
self-referential standard survives contact with experiment is the question
the rest of the paper answers.

\section{Pre-registered protocol}
\label{sec:protocol}

The validation question was fixed as: can the active GNN predict Na-ion
average voltages well enough to drive screening decisions? Before computing
any bias-corrected metric we committed a verdict ladder, a primary metric, and
four failure modes (F1--F4) to the repository; a fifth stop condition (F5) was
fixed in the Step-3 protocol before literature curation began. The committed
documents, their hashes, and the verbatim definitions are reproduced in
Appendix~\ref{app:failure}.

The ladder maps the held-out mean absolute error (MAE) to one of three
verdicts: screening-grade below \SI{0.30}{\volt}, ranking-only between
\SI{0.30}{\volt} and \SI{0.50}{\volt}, and not screening-grade above
\SI{0.50}{\volt}. The primary metric is deliberately conservative: the upper
edge of the 95\% bootstrap confidence interval on the held-out, bias-corrected
MAE, where ``held-out'' requires that no compound's error contributed to the
bias estimate applied to it. Point estimates, in-sample numbers, and
best-family numbers are reported but cannot set the verdict.

The failure modes encode the ways a calibration claim could be technically
true and practically meaningless. F1 fires when per-family bias estimates
spread by more than \SI{0.15}{\volt}, in which case a single additive
correction does not generalize and no clean verdict is issued. F2 fires on
any in-sample contamination of the bias estimate. F3 marks any verdict from
fewer than twenty held-out compounds as provisional. F4 fires when family
biases disagree in sign. F5 is a curation stop: if more than half of the
compounds in the experimental upgrade cannot be cited from primary literature,
the subset is not defensibly experimental, and the analysis halts rather than
estimating its way to a sample. The pre-registration also commits revision
triggers that outlast publication: among them, two in-box compounds
disagreeing with corrected predictions by more than \SI{0.4}{\volt} retire the
bias-corrected predictor entirely.

Reference curation followed a quote-anchored discipline. Every literature
voltage carries a verbatim snippet from a source fetched during the run,
a DOI, the cell configuration, and the reference electrode; values against
anything other than Na metal were rejected. Each datum was assigned an
evidence grade, defined at curation time and not pre-registered, of A
(average voltage stated by the authors in fetched text), B (single arithmetic
step from author-stated plateaus or phase windows, calculation recorded), or C
(read off a published figure; low confidence). Grade D (no anchorable source)
meant the compound was dropped, and five compounds were dropped rather than
estimated (Appendix~\ref{app:curation}). Every polymorph was lattice-verified
against the MP structure the GNN actually ingested, because the screen
predicts on a specific crystal, not on a formula. Finally, the curated table
went to a human audit: the operator manually verified the five
most verdict-sensitive extractions against the primary sources, all five of
which were confirmed, and that audit reclassified one row (maricite \ce{NaFePO4})
as a phase-identity mismatch, which defines the canonical $n=6$ metric set
used below.

\section{Results A: the screen against references, then against experiment}
\label{sec:resultsA}

\subsection{Against computed references}
\label{sec:resultsA1}

The held-out test that motivated everything else used 74 Na-ion cathode
entries from the MP battery database whose discharged structures do not
appear in the training corpus, with the MP computed average voltage as the
reference. The raw held-out MAE was \SI{0.5892}{\volt} with a mean signed
error of \SI{+0.4263}{\volt} and a Pearson correlation between prediction and
reference of $0.28$. Most of the error, in other words, looked like a single
large positive bias, which is exactly the situation an additive correction
is supposed to repair, and exactly where the pre-registered machinery earned
its keep.

\begin{table}[tb]
\caption{\label{tab:family}Per-family raw bias and MAE of the active GNN
against MP computed reference voltages on the 74-compound held-out set.
The spread between family biases (families with $n\ge5$) is
\SI{0.4419}{\volt}, which fires pre-registered failure mode F1: no single
additive correction generalizes across families.}
\input{tables/t3_family_bias}
\end{table}

The bias is not one number. Grouped by chemistry family
(Table~\ref{tab:family}), the per-family signed errors span
\SI{0.4419}{\volt} between the mixed-anion ``other'' bucket
(\SI{+0.6316}{\volt}, $n=38$) and the layered oxides (\SI{+0.1897}{\volt},
$n=9$), with the polyanionic phosphates (the family the Na campaign
targeted) at \SI{+0.2022}{\volt} ($n=26$). That spread exceeds the
pre-registered \SI{0.15}{\volt} threshold, so F1 fired and no clean verdict
tier could be issued. The leave-one-family-out estimate of what an additive
correction would actually deliver made the point mechanically: the corrected
held-out MAE ($n=64$) was \SI{0.5644}{\volt}, \emph{worse-behaved} than it
appears since the residual mean bias of \SI{+0.1101}{\volt} should have been
near zero if the correction generalized, and the conservative upper
95\%-confidence value was \SI{0.6610}{\volt}. Cross-family bias transfer
moves error around instead of removing it.

These numbers compare the model to MP's computed voltages, so they bound the
model's agreement with its own reference scale, not with electrochemistry.
The pre-registered plan was therefore to upgrade the reference on the
phosphate subset to literature experimental values. That upgrade failed in a
way we did not anticipate and now consider a finding in its own right: of the
26 held-out polyanionic phosphates, the number with a citable experimental
average voltage in primary literature was zero. Eighteen of the 26 have no
Na-cell publication of any kind; the remainder appear only in structural
crystallography or report cycling windows without an extractable average.
The stop condition F5 (threshold: half the subset uncitable) fired at 100\%,
and the analysis halted rather than substituting estimates. The held-out
set, drawn from a computed battery database, was not a set of materials
anyone had measured. A reference upgrade therefore required building a new
compound set from the experimental literature inward, which is Step 3b.

\subsection{Against experiment}
\label{sec:resultsA2}

\begin{figure}[tb]
\includegraphics[width=\columnwidth]{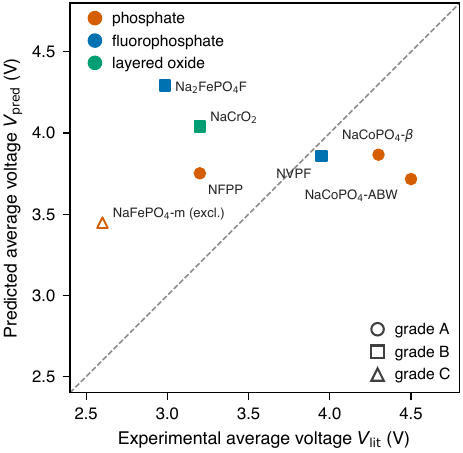}
\caption{\label{fig:parity}Active-model predictions against quote-anchored
experimental average voltages for the curated Na-ion cathode set, colored by
chemistry family with evidence grades marked. The open symbol is maricite
\ce{NaFePO4}, excluded from the canonical metrics after the operator audit
identified a phase-identity mismatch (the cycled phase is amorphous, the
predicted structure crystalline). The diagonal is parity. Predictions
compress into a band near 3.4--4.3~V across references spanning
2.6--4.5~V.}
\end{figure}

The curated set contains nine literature-validated rows: five evidence grade
A, three grade B, one grade C, spanning phosphates, fluorophosphates, one
sulfate, and one layered oxide (Appendix~\ref{app:curation},
Table~\ref{tab:curated}). Two rows (\ce{Na2FeP2O7} and alluaudite
\ce{Na2Fe2(SO4)3}) have no MP structure for the GNN to ingest and are
excluded from metrics; seven were predicted. The operator audit then
excluded maricite \ce{NaFePO4} for phase identity: its
\SI{2.6}{\volt} literature value belongs to the amorphous phase formed on
first desodiation \cite{kim2015}, while the model predicted on crystalline
maricite, leaving $n=6$ as the canonical set. Every verdict below is
unchanged at $n=7$.

\begin{table*}[tb]
\caption{\label{tab:metrics}Held-out error of the active GNN against
experimental average voltages, for all computed variants. The pre-registered
primary metric is the upper edge of the 95\% bootstrap confidence interval
(CI) on the bias-corrected MAE; every variant of every estimate, including
every lower CI edge of the corrected metric, exceeds the \SI{0.50}{\volt}
``not screening-grade'' threshold.}
\input{tables/t2_metrics}
\end{table*}

\begin{figure}[tb]
\includegraphics[width=\columnwidth]{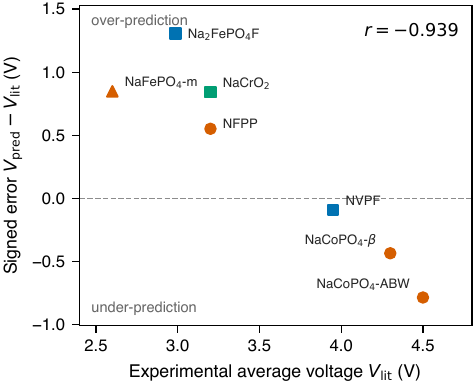}
\caption{\label{fig:residuals}Signed prediction error against the
experimental reference voltage for the seven predicted rows. The correlation
$r=-0.939$ quantifies the compression: over-prediction below
$\sim$\SI{3.5}{\volt}, under-prediction of the \SIrange{4.3}{4.5}{\volt}
cobalt rows. No additive correction can remove a residual with this
structure, which is why the pre-registered correction worsens the MAE
(Table~\ref{tab:metrics}).}
\end{figure}

The raw held-out MAE on the canonical set is \SI{0.668}{\volt} with a mean
signed error of \SI{+0.231}{\volt} (Table~\ref{tab:metrics}). Applying the
pre-registered leave-one-out additive bias correction makes the error larger,
not smaller: \SI{0.802}{\volt} corrected, with an upper 95\% confidence edge
of \SI{1.092}{\volt}, which is the pre-registered primary metric. The
mechanism is visible in Fig.~\ref{fig:residuals}: the signed error is almost
a linear function of the reference voltage, with Pearson
$r(\mathrm{err},\vlit)=-0.939$. The model over-predicts every compound below
about \SI{3.5}{\volt} (maricite \SI{+0.85}{\volt}, NFPP \SI{+0.55}{\volt},
\ce{Na2FePO4F} \SI{+1.31}{\volt}, \ce{NaCrO2} \SI{+0.84}{\volt}) and
under-predicts the two high-voltage cobalt polymorphs (\SI{-0.78}{\volt} and
\SI{-0.43}{\volt}), compressing a \SI{1.9}{\volt} experimental range into
roughly half that span of predictions. A single additive shift cannot repair
a sign-flipping, voltage-dependent residual; subtracting the mean bias
improves the low-voltage rows while pushing the cobalt rows further from the
measurements. This is the experimental confirmation of the F1 structure seen
against computed references in Sec.~\ref{sec:resultsA1}.

Against the pre-registered ladder, the conservative primary metric of
\SI{1.092}{\volt} sits far above the \SI{0.50}{\volt} threshold: the screen
is \emph{not screening-grade}, with the verdict marked provisional because F3
fired ($n<20$). The verdict needs no benefit of the doubt: the raw point
estimate, the corrected point estimate, and both confidence edges of every
computed variant (with and without the grade-C row; full set and phosphate
subset) all individually exceed \SI{0.50}{\volt}
(Table~\ref{tab:metrics}). The pre-registered revision trigger also engaged:
two in-box iron phosphates miss by more than the \SI{0.4}{\volt} trigger
threshold in raw error, which retires the bias-corrected predictor outright.
As a reproducibility check, the two rows shared with the
Sec.~\ref{sec:resultsA1} analysis reproduce their earlier predictions within
Monte-Carlo-dropout noise ($\Delta=\SI{-0.011}{\volt}$ and
$\SI{-0.055}{\volt}$). The operator audit confirmed all five spot-checked
extractions, so the failure cannot be attributed to careless curation. The
screen, evaluated exactly as we committed to evaluate it, does not work.

\section{Results B: the reference scale and the prior-art base rate}
\label{sec:resultsB}

\subsection{Decomposing prediction error against the reference}
\label{sec:resultsB1}

\begin{figure}[tb]
\includegraphics[width=\columnwidth]{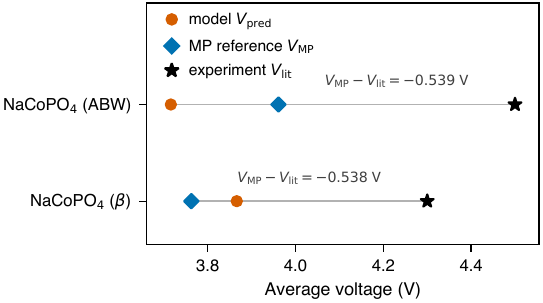}
\caption{\label{fig:decomp}Three-way comparison for the two
polymorph-resolved \ce{NaCoPO4} rows: model prediction, MP computed
reference, and quote-anchored experimental value. The reference sits
$\approx$\SI{0.54}{\volt} below measurement on both polymorphs; the
model--reference difference is smaller and changes sign.}
\end{figure}

Two curated rows, the ABW and $\beta$ polymorphs of \ce{NaCoPO4}, also
appear in the MP battery database with computed average voltages, which
permits the three-way decomposition
$(\vpred-\vlit) = (\vpred-\vmp) + (\vmp-\vlit)$ on polymorph-resolved
structures. The result is lopsided (Fig.~\ref{fig:decomp}). The MP PBE+$U$
reference sits \SI{0.539}{\volt} below the measured value for the ABW
polymorph and \SI{0.538}{\volt} below it for $\beta$, the same offset
twice, while the model's deviation from its own reference is smaller and
mixed in sign ($-0.245$ and \SI{+0.104}{\volt}). For the $\beta$ polymorph
the model lands \emph{closer} to experiment than the reference it was
trained toward. On these rows, most of the prediction-versus-experiment gap
is reference error, not model error.

Two caveats bound this finding, and we state them at full strength. First,
$n=2$: this is a direction, not an estimate. Second, the \ce{NaCoPO4}
literature values are first-charge averages from cells with roughly 11\%
first-cycle reversibility \cite{chiring2021}, and kinetic polarization
inflates a first-charge average above the equilibrium voltage, so part of
the \SI{0.54}{\volt} is experimental upper bias rather than DFT error. What
makes the direction credible despite $n=2$ is independent replication on our
own bench: our PBE+$U$ calculation of \ce{Na3V2(PO4)3} (a different
compound, a different code installation, an empirically fitted $U_V$) gives
\SI{2.8957}{\volt} against a measured plateau near \SI{3.4}{\volt}
\cite{jian2012}, an offset of the same sign and similar size. Published
GGA+$U$ benchmarking reports underestimation of transition-metal redox
voltages of comparable scale when $U$ is not tuned per couple
\cite{wang2006,aydinol1997}. Three independent comparisons agree in
direction; none alone is conclusive.

The consequence reaches beyond our pipeline. The Sec.~\ref{sec:resultsA1}
family analysis found the model predicting \SI{+0.20}{\volt} \emph{above} MP
references on phosphates, which looked like model bias; against experiment,
those same predictions are too low. A model trained toward MP voltages in
this chemistry inherits a deficit of roughly \SIrange{0.4}{0.5}{\volt}
against measurement, and the better it fits its references the more
faithfully it reproduces their offset. Any screen whose accuracy statement
ends at ``versus MP'' should be read with that substitution in mind.

\subsection{Reference integrity as a finding}
\label{sec:resultsB2}

Building the curated set surfaced a class of errors that no accuracy metric
captures: identity errors in the reference chain. Machine validation of the
initial human-drafted curation proposal found six of its MP identifiers
wrong. One was a namespace collision: the battery-database identifier
matching \ce{Na4Fe3(PO4)2(P2O7)} actually keys a \ce{Cr3O8} battery pair, so
a training-set membership flag based on identifier matching was simply false.
The others pointed at unrelated compounds (\ce{Fe6O5F7},
\ce{CaWO4}, \ce{Li3MnCr3O8}, \ce{Na7(CoO3)2}) or did not exist. Phase
identity was the other recurring trap. The maricite \ce{NaFePO4} literature
voltage belongs to an amorphous phase created by the first charge
\cite{kim2015}; the MP entry alleged to be maricite \ce{NaMnPO4} is a
different compound (\ce{Na2Mn3(PO4)3}); and the literature identity of
\ce{NaVPO4F} is itself disputed between a tavorite phase and
\ce{Na3V2(PO4)2F3} \cite{barker2003}. Each of these, uncaught, silently
corrupts a validation pair: the model predicts on one crystal while the
reference describes another. The full correction and drop log is
Appendix~\ref{app:curation}. We draw a methodological conclusion:
in this regime, reference-data integrity, not model architecture, is the
binding constraint on what a validation can even mean.

\subsection{Base rates in the targeted space}
\label{sec:resultsB3}

A screen also cannot be useful if the space it screens contains nothing left
to find. Enumerating the generator's own targeted space (earth-abundant
Na cathode compositions over its standard framework library) gives 390
distinct formulas, of which 78 survive charge-neutrality and
electronegativity screening. The prior-art layer, which can only downgrade,
flags 55 of those 78 (70.5\%) as already published: 52 by structural-family
match, 20 by exact composition in the Crystallography Open Database, 2 by
composition-verified literature citation (paths overlap). Because the layer
certifies nothing and two of its sources are stubs, 70.5\% is a lower bound.
The 23 surviving formulas are substitutional variants of two known families,
and at least one survivor is a known cathode the title-level checker missed,
a measured precision disclosure rather than a footnote. The targeted space is
saturated in exactly the sense Cheetham and Seshadri describe for generated
libraries at large \cite{cheetham2024}.

\begin{table}[tb]
\caption{\label{tab:baserate}Prior-art base-rate probes. Top: the
generator's targeted earth-abundant Na space. Bottom: a pinned uniform
sample of the GNoME stable-materials release screened with the same layer,
and the staged triage funnel applied to the prior-art-absent remainder.
Flagged rates are lower bounds (the layer only downgrades); the GNoME
verified floor reflects a hand audit of the five flagged hits (3/5
confirmed). The stability disputes are a disagreement rate between two
models, not a correctness measure.}
\input{tables/t5_baserate}
\end{table}

To test whether that base-rate problem is ours alone, we ran the same layer
over a pinned uniform random sample ($N=500$, fixed seed and identifier
list) of the 554{,}054-entry GNoME stable-materials release
\cite{merchant2023}, followed by our staged triage funnel on the 495
prior-art-absent survivors (Table~\ref{tab:baserate}). The title-level
screen flagged 1.0\% of the sample; hand-auditing those five hits confirmed
three, a verified floor of 0.6\%, and an exact-composition check against the
Crystallography Open Database matched 0 of 500. Read correctly, these
numbers say the checkable-known fraction of GNoME under \emph{this} layer's
coverage is small: they measure our coverage as much as GNoME's novelty,
and they are consistent with the published finding that GNoME's overlap with
the known literature is concentrated in ways title-level matching does not
see \cite{cheetham2024,leeman2024}. The funnel then removed 460 of the 495
survivors (SMACT validity 446, stability screen 154, with overlap; one
polaron block), leaving 35 (7.1\%). Our borrowed stability model disputes
the GNoME stable label on 31.1\% of the sample; we report that as a
disagreement rate between two models and make no claim about which is right.
None of the 35 funnel survivors pairs a favorable sustainability profile
with a checkable application fit. A uniform sample of the flagship generated
library, passed through this stack, yields no candidate this pipeline could
defensibly act on, which is a statement about the pipeline and the
library jointly, and a base rate any triage system in this field has to
beat.

\section{Results C: calibrating our own bench}
\label{sec:resultsC}

Sections~\ref{sec:resultsA} and \ref{sec:resultsB} indict the reference
scale we and others train against; consistency requires applying the same
standard to our own DFT bench. We pre-registered and ran an offset audit of
the QME PBE+$U$ scale against experimental plateaus on known Li-ion couples.
The design was committed and the decision rule fixed before any run; the
result is reported below.

\begin{table}[tb]
\caption{\label{tab:d1}The pre-registered offset audit. Each couple is
computed as a paired vc-relax on the local bench under the anchor-scale
settings (Sec.~\ref{sec:methods}); $\delta = V_{\mathrm{QME}} -
V_{\mathrm{exp}}$. The \ce{LiMn2O4} pair is a pre-registered stretch
compound: if either half converges outside its registered spin manifold,
the verdict is issued on the $n=3$ core and the exclusion is reported. The
verdict gate, fixed before any run: $\mathrm{sd}(\delta) < \SI{0.15}{\volt}$
means the scale is calibratable and voltages publish with the calibrated
offset and error bars; $\mathrm{sd}(\delta) \ge \SI{0.3}{\volt}$ retires
absolute-voltage claims pipeline-wide in favor of ranking-only language; the
zone between is an operator decision documented at sync time.}
\input{tables/t4_d1design}
\end{table}

The audit computes the average voltage of four couples with unimpeachable
experimental plateaus at the same redox couples the bench would compute:
\ce{LiCoO2 -> Li_{0.5}CoO2} (existing anchor, measured window
\SIrange{3.9}{4.1}{\volt}), \ce{LiFePO4 -> FePO4}
(\SI{3.45}{\volt} \cite{padhi1997}), \ce{Li2FeP2O7 -> LiFeP2O7}
(\SI{3.5}{\volt} \cite{nishimura2010}), and \ce{LiMn2O4} $\to$
$\lambda$-\ce{MnO2}
(\SI{4.05}{\volt} \cite{thackeray1983,ohzuku1990}), the last as a stretch
compound whose exclusion path was committed in advance because its true low-temperature
ground state is charge-ordered \cite{rodriguez1998} and the
ferromagnetic-cubic convention adds a documented model-choice uncertainty
(Table~\ref{tab:d1}). Lithium counts per cell are asserted from the staged
structures, not formulas, and the committed analysis script refuses to
produce a verdict on any mismatch or on an incomplete job set.

The audit is complete on all four couples, each relaxed under a converged
plane-wave cutoff (Methods). The offsets are $\delta=+0.31$, $+0.48$, $+0.43$,
and $-\SI{0.20}{\volt}$ for \ce{LiCoO2}, \ce{LiFePO4}, \ce{Li2FeP2O7}, and
\ce{LiMn2O4} (Table~\ref{tab:d1}), with $\mathrm{sd}(\delta)=\SI{0.31}{\volt}$
over the $n=4$ set, above the pre-registered \SI{0.30}{\volt} bar. The gate
fires as committed: absolute-voltage language is retired from our own DFT
ledger in favor of ranking-only claims. The failure is chemistry, not noise.
The three non-stretch couples span two redox metals and three structure types
(layered \ce{Co^3+/4+}, olivine and pyrophosphate \ce{Fe^2+/3+}) yet agree to
\SI{0.09}{\volt}, a common offset of $+\SI{0.41}{\volt}$. The outlier is the
\ce{Mn} spinel stretch ($-0.20$), pre-registered as model-choice-uncertain
because its charge-ordered ground state is not captured by the
ferromagnetic-cubic convention; it converged within its registered spin
manifold, so the committed gate retains it, and its inclusion is what carries
the set over threshold. A single global offset is therefore rejected while a
per-chemistry offset is admissible. Two notes belong with these numbers. The
offsets are meaningful only at a converged charge-density cutoff: at the
\SI{200}{\rydberg} anchor-scale the \ce{LiFePO4} cell relaxed to a spuriously
large volume that inflated its voltage by \SI{0.6}{\volt}, the \ce{Li2FeP2O7}
relaxation did not converge at all, and \ce{LiCoO2} itself shifted
\SI{0.19}{\volt}; all four couples are reported at \SI{560}{\rydberg}
(\ce{LiCoO2} verified stable to \SI{4}{\milli\volt} at \SI{720}{\rydberg}).
This cutoff change postdates the pre-registration, so we state plainly that
the gate outcome does not depend on it: at the pre-registered \SI{200}{\rydberg}
cutoff the evaluable couples already disperse with $\mathrm{sd}(\delta)\approx
\SI{0.68}{\volt}$, more than twice the bar, so the audit fails at both cutoffs
and the converged recomputation only sharpens the magnitudes.
The \ce{Li2FeP2O7} and \ce{LiFeP2O7} endpoints are energy-converged but
force-plateaued in a soft basin (residual force below
\SI{0.006}{\rydberg\per\bohr}), accepted with disclosure, with spin manifolds
matching the registered ground states.

\section{Discussion}
\label{sec:discussion}

The findings travel beyond this pipeline in proportion to how common its
ingredients are, and its ingredients are the field's defaults: a GNN
property screen trained where data is plentiful and deployed where it is
not, computed reference voltages standing in for measurements, and a novelty
notion defined by absence from the databases at hand. On the evidence here,
each default fails in a specific, measurable way. The screen's residual is
structured, not noisy, so the standard additive recalibration is not merely
insufficient but counterproductive: it moves error between voltage
regimes. The reference scale is offset from experiment by about half a volt
in this chemistry, in the same direction on every row we could decompose, so
benchmark MAEs quoted against computed references understate true error for
exactly the high-voltage chemistries where screening matters most. And the
base rate of already-published compositions in a targeted substitution space
is high enough ($\ge$70\%) that a screen with perfect accuracy would still
mostly re-rank known materials.

We are deliberate about what the decomposition result does not show. It
does not show that MP voltages are wrong by \SI{0.54}{\volt} in general;
$n=2$, one chemistry, first-charge references. It shows that on the only
rows where the comparison was possible at polymorph resolution, the
reference term dominated the model term, and that our own independent
PBE+$U$ calculation reproduces the offset's direction and rough size on a
third compound. The honest generalization is conditional: models trained
toward GGA+$U$ intercalation voltages should be presumed to inherit a
systematic deficit against experiment in transition-metal phosphate
chemistries until a per-chemistry offset audit shows otherwise. That
presumption is cheap to test and expensive to ignore.

The novelty result has a constructive reading. Because the prior-art layer
is precision-first and downgrade-only, its verdict for any surviving
candidate is not ``novel'' but ``no prior art found within the stated
coverage'': named sources, named match rules, a dated coverage statement,
and a hand-measured precision on known and fictional controls. We think
this bounded-coverage absence claim is the strongest novelty statement an
automated system should ever issue. The alternative, treating absence
from a convenient corpus as discovery, is how a field ends up
re-announcing 1980s solid-state chemistry, and our own targeted space was
70\% re-announcement before the screen ever ran. Subscription structure
databases we lack (notably ICSD) would raise coverage, and we state plainly
that a zero-budget instantiation cannot close that gap; it can only declare
it.

The strongest objection we know how to state is this: the entire exercise is
self-referential: a five-anchor ledger of voltage-only results on known
compounds, four of them at couples no electrolyte can cycle, gated by a
model trained on those same anchors, validated by the same operator who
built the system, on six experimental points partly contaminated by
first-charge kinetics; nothing here is independently verified, so why should
the negative result be trusted more than the system it indicts? Our answer
is the protocol, not the pipeline. The thresholds, metric, and failure
modes were committed before the metrics existed and are quotable from the
repository history; the conservative metric was designed against our own
favor; the curation is quote-anchored to fetched sources with a per-row
audit trail; an independent human audit of the most verdict-sensitive rows
confirmed five of five extractions and \emph{strengthened} the negative
verdict by excluding a row that flattered the model; and every number in
this paper, including the unflattering ones, resolves to a committed
artifact. A skeptic who distrusts our system should distrust it
symmetrically: the same self-referential gates it passed (anchor-fit,
held-out cross-validation on its own DFT) are the gates we are arguing the
field should stop treating as validation. The negative result survives the
objection because the objection \emph{is} the result.

Several limitations bound all of the above and are conceded rather than
argued. The experimental comparison rests on $n=6$ ($n=7$ including the
grade-C row; verdict unchanged), the decomposition on $n=2$, and no
bootstrap can manufacture sample size that curation could not. One
reference (\ce{NaCrO2}) is a capacity-weighted scalar we computed from
author-stated phase windows; plausible protocol choices move it by
\SIrange{0.2}{0.3}{\volt} in the direction that worsens, not improves, the
model's residual. The anchor ledger mixes runs from two Quantum ESPRESSO
installations (a retired cloud host and the local bench) under identical
cutoffs, pseudopotentials, and $U$ values, with per-run provenance recorded;
the offset audit (Sec.~\ref{sec:resultsC}) re-computed four couples on the
local scale at a converged cutoff, finding the anchor-scale density cutoff
itself under-converged. The stability and embedding model is borrowed,
and its 31.1\% disagreement with GNoME labels is unvalidated in both
directions. The acquisition stage is unevaluated and unevaluable at $N=5$.
The pipeline has never verified a candidate absent from its corpora, so its
discovery half is untested by construction. And the operator-validation
step, while documented verbatim in the repository, was performed by the
system's author; a third-party replication of the curation from the
committed CSV is the cheapest external check this work admits and is
explicitly invited.

What survives is narrow but load-bearing: a validation protocol that cannot
be renegotiated after the fact, a quote-anchored experimental reference set
with full provenance for a chemistry where such sets are scarce, evidence
that computed voltage references carry a directional, half-volt-scale error
into everything trained on them, and one pipeline that now says ``not
screening-grade'' about itself in public, with the artifacts to back it.
We retire the Na screen, we do not retrain it on more computed voltages,
and our own bench's absolute-voltage language now hangs on a pre-registered
audit it may fail. That is what we believe validation discipline looks like
when the answer is no.

\section{Methods}
\label{sec:methods}

\subsection{DFT verification bench}
All verification runs are spin-polarized PBE+$U$ variable-cell relaxations
in Quantum ESPRESSO \cite{giannozzi2009,giannozzi2017}, with wavefunction
and charge-density cutoffs of 50~Ry and 200~Ry, Gaussian
smearing of 0.01~Ry, local Thomas--Fermi mixing, and Monkhorst--Pack
$k$-grids \cite{monkhorst1976} set per cell by the input stager and recorded
with each run (the anchor cells used $3\times3\times3$; the audit cells use
denser grids, e.g.\ $6\times5\times3$ for olivine \ce{LiFePO4}). Hubbard
corrections use the Dudarev formulation \cite{dudarev1998}: on the current
QE~7.3.1 bench through the \texttt{HUBBARD \{ortho-atomic\}} card, and on
the retired QE~6.4.x cloud host through the legacy \texttt{lda\_plus\_u}
interface; $U_{\mathrm{eff}}$ values are Fe 5.3~eV, Co 3.4~eV,
Ni 6.2~eV, Mn 3.9~eV (literature-standard \cite{wang2006,jain2013}),
and V 3.1~eV (an empirical fit, recorded as such per run). No
first-principles $U$ exists anywhere in the pipeline: one linear-response
attempt \cite{timrov2022} failed to converge and was not repeated, and we
flag every $U$ provenance in the database rather than claim otherwise.
Pseudopotentials are pinned by file name (Appendix~\ref{app:ledger}); the
backbone sets are GBRV ultrasoft \cite{garrity2014} and PSlibrary PAW/USPP
\cite{dalcorso2014}.

The average intercalation voltage of a couple with $n$ cycling ions follows
the standard total-energy construction \cite{aydinol1997},
\begin{equation}
\label{eq:voltage}
V = -\frac{E_{\mathrm{charged+ion}} - E_{\mathrm{charged}} - n\,\mu_{\mathrm{ion}}}
{n}\,,
\end{equation}
with energies in Ry converted at 13.605693122994~eV/Ry,
$\mu_{\mathrm{Li}} = -14.4725646547$~Ry from a BCC Li reference run
and $\mu_{\mathrm{Na}} = -95.34612073$~Ry from the analogous Na
reference, both stored in the run database. Ion counts $n$ are counted from
the relaxed cells, never inferred from formulas, and are hard-asserted by
the analysis scripts. The parsed energy is the BFGS \texttt{Final enthalpy};
the post-convergence single-point rewrite is never used. Spin assignment is
pre-registered per run in the ferromagnetic convention standard for GGA+$U$
voltage work \cite{wang2006}, with the expected high-spin manifolds and
total magnetizations fixed in advance; a pair whose halves converge in
different manifolds is excluded rather than mixed, and one ledger row
(\ce{Na3Fe2(PO4)3}, desodiated) converged only under a fixed total
magnetization, which is recorded with the run. Calculations executed
sequentially on a Mac Mini M4 (local bench) and, for the early anchor runs,
on a retired cloud host; the backend of every run is a database field, and
identical cutoffs, pseudopotentials, and $U$ values were used on both.

\subsection{Screen, training data, and gates}
The voltage screen is a Siamese multi-head GNN over (charged, discharged)
crystal-graph pairs with Monte-Carlo dropout for predictive uncertainty.
Its training corpus holds 2814 graphs assembled from MP battery pairs, of
which 20 pairs are Na chemistries. Candidate relaxation during screening
uses M3GNet \cite{chen2022}; stability estimates and the 128-dimensional
embedding space come from MACE-MP-0 \cite{batatia2022,batatia2024} with
property heads. The deployed model is the first PBE+$U$ fine-tune of the
v2.4 baseline, trained against the five anchor voltages with elevated
anchor weights; activation was gated on the fitted anchor error
(\SI{0.084}{\volt} mean absolute) and on a held-out leave-one-out
cross-validation across the anchors. We report the cross-validation gate
qualitatively because its fold-level outputs were not committed to the
repository, and this paper does not quote numbers without committed
artifacts. Both gates measure agreement with the pipeline's own DFT scale.

\subsection{Validation statistics}
The Sec.~\ref{sec:resultsA1} analysis holds out 74 MP battery entries whose
discharged identifiers are absent from the training corpus, with
family-restricted leave-one-family-out bias correction over families with at
least ten members and a 10{,}000-resample bootstrap \cite{efron1979}
(seed 42) for confidence intervals. The Sec.~\ref{sec:resultsA2} analysis uses leave-one-out additive
bias correction, as pre-registered for $5\le n<10$, with a 10{,}000-resample
bootstrap (seed 20260609); all seven predicted rows are held out by
construction (none is in the training corpus, verified by composition
mapping rather than identifier string match after the collision of
Sec.~\ref{sec:resultsB2}). The primary metric, fixed before computation, is
the upper edge of the 95\% bootstrap interval on the bias-corrected held-out
MAE.

\subsection{Curation and audit}
Literature values were extracted only from sources fetched during the
curation run, with the supporting sentence or arithmetic recorded per row,
the reference electrode checked to be Na metal, and the polymorph
lattice-verified against the ingested MP structure. Evidence grades A--C are
defined in Sec.~\ref{sec:protocol}; grade D rows were dropped. The
operator audit verified the five most verdict-sensitive rows against primary
sources (five confirmed) and excluded the maricite row for phase identity;
the audit record, including one clerical heading error, is carried verbatim
in the repository and summarized in Appendix~\ref{app:curation}.

\subsection{Prior-art layer and base-rate probes}
The layer searches OpenAlex and Crossref at title level with
composition-verified matching (exact, de-intercalated framework, or element
set), an offline structural-family screen, and exact reduced-composition
lookup against the Crystallography Open Database via OPTIMADE; patent and
ICSD sources are declared stubs that fail open. Its precision audit uses
8 known cathodes and 5 fictional compositions. The GNoME probe draws a
uniform random sample ($N=500$, seed 42, identifier list pinned in the
repository) from the SHA-pinned public stable-materials summary (554{,}054
rows), screens it with the unmodified layer, hand-audits every flagged hit,
and then applies the staged funnel (SMACT validity \cite{davies2019},
borrowed-model stability, $E_f<0$ and hull distance
$<0.15$~eV/atom, and a non-blocking polaron screen) to the
prior-art-absent remainder, using the GNoME-relaxed structures from the
public release. All probe inputs, outputs, and seeds are committed.

\begin{acknowledgments}
Manuscript drafting, figure and analysis tooling, and literature-metadata
verification were assisted by Anthropic's Claude (Fable 5); all quantitative
claims were generated from, and verified against, the committed repository
artifacts, and the author takes sole responsibility for the content.
\end{acknowledgments}

\paragraph*{Data availability.}
All committed artifacts cited in this paper, namely the pre-registration
documents, the curated reference set with quote anchors, the per-row
results, the analysis scripts, and the figure-generation code, are
available at \url{https://github.com/Krishnatejavepa/qme-paper-validation}
at the tagged commit; the ancillary files of this submission mirror the
validation set and analysis scripts.

\appendix

\section{Curated experimental reference set}
\label{app:curated}

Table~\ref{tab:curated} lists the nine literature-validated rows with
polymorph, evidence grade, reference voltage, and DOI. Rows without an MP
structure are excluded from metrics. The \ce{NaCrO2} scalar is the
capacity-weighted average of the author-stated phase windows,
$(0.25\cdot2.85+0.35\cdot3.43)/0.60=\SI{3.19}{\volt}\to\SI{3.20}{\volt}$
\cite{bo2016}; common fixed-window cycling protocols
(\SIrange{2.5}{3.6}{\volt}) yield \SIrange{2.9}{3.0}{\volt}, which would
increase the row's $+0.84$~V residual and leave every verdict unchanged.
The two \ce{NaCoPO4} values are first-charge averages
(Sec.~\ref{sec:resultsB1}); the \SI{0.1}{\volt} internal spread between the
abstract and figure readings of the $\beta$ polymorph is recorded in the
curation file. Primary sources for the rows not already cited in the text:
\ce{Na4Fe3(PO4)2(P2O7)} \cite{kim2012}, \ce{Na3V2(PO4)2F3}
\cite{yan2019}, \ce{Na2FeP2O7} \cite{barpanda2012}, and alluaudite
\ce{Na2Fe2(SO4)3} \cite{barpanda2014}.

\begin{table*}[tb]
\caption{\label{tab:curated}The curated Na-ion cathode reference set.
Grade: A = author-stated average in fetched text; B = one arithmetic step
from author-stated plateaus or windows (calculation recorded); C = figure
read-off (low confidence). The maricite row is excluded from canonical
metrics by the operator audit (phase identity: the cycled phase is
amorphous). Rows marked ``no MP entry'' are literature-validated but cannot
be predicted by the structure-ingesting screen.}
\input{tables/t1_curated}
\end{table*}

\section{Curation log: corrections, drops, and the audit record}
\label{app:curation}

Machine validation corrected six MP identifiers from the human curation
proposal: \ce{NaFePO4} mp-755097$\to$mp-19226 (the former is
\ce{Fe6O5F7}); \ce{Na2FeP2O7} mp-19426$\to$no MP entry (the former is
\ce{CaWO4}); \ce{Na3V2(PO4)2F3} mp-755519$\to$mp-694937 (the former is
\ce{Li3MnCr3O8}); \ce{NaCrO2} mp-19427$\to$mp-578604 (the former is
\ce{Na7(CoO3)2}); \ce{Na2FePO4F} mp-22162$\to$mp-1194940 (the former does
not resolve); and the in-training flag for \ce{Na4Fe3(PO4)2(P2O7)} was a
battery-identifier namespace collision with a \ce{Cr3O8} pair. Five
compounds were dropped rather than estimated: maricite \ce{NaMnPO4} (no
defensible average voltage exists; the best-enabled material reaches
$\sim$30\% of theoretical capacity with \SI{2.4}{\volt} hysteresis
\cite{mohsin2023}, and maricite inactivity is independently reported
\cite{chiring2021}); \ce{NaVPO4F} (full-cell-versus-hard-carbon voltage,
disputed phase identity, and an MP entry matching neither claimed phase
\cite{barker2003}); \ce{NaFeSO4F} (synthesis and structure literature only
\cite{tripathi2010}); \ce{NaNi_{0.5}Mn_{0.5}O2} (window stated without an
average and no ingestible MP structure \cite{komaba2012}); and
P2-\ce{Na_{2/3}MnO2} (no matching MP structure; structural literature
only). The proposal's \ce{Na2FePO4F} value also moved from the Li-cell
number \SI{3.5}{\volt} \cite{ellis2007} to the Na-cell plateaus of Kawabe
\emph{et al.} \cite{kawabe2011}, averaged to \SI{2.985}{\volt}.

The operator validation record is carried verbatim in the repository file
\texttt{STEP3B\_OPERATOR\_VALIDATION\_2026-06-09.md}. Two of its features
are preserved rather than edited, per the carried-record rule: its Section~2
heading reads ``\ce{Na2FeP2O7}'' over content that describes the
\ce{Na2FePO4F} row (a clerical error flagged in the file's machine note),
and its Section~3 illustrates the post-hoc nature of any \ce{NaCrO2} scalar
with the midpoint arithmetic $(3.1+3.7)/2=\SI{3.4}{\volt}$; the canonical
value used throughout this paper is the capacity-weighted \SI{3.20}{\volt}
of Appendix~\ref{app:curated}, and the illustration's existence is part of
why that row carries grade B rather than A.

In the earlier Step-3 attempt on the 74-compound held-out set, 0 of the 26
polyanionic phosphates had a citable experimental average voltage (the F5
stop condition fired at 100\%, against a 50\% threshold): 18 had no Na-cell
publication of any kind, 3 had structural crystallography only, 2 matched
only related-but-different compounds, and 3 (the \ce{NaCoPO4} polymorphs and
maricite \ce{NaMnPO4}) had cathode literature without an extractable
average. The per-compound search log is the committed file
\texttt{step3\_curation\_log.csv}.

\section{Pre-registered definitions, verbatim}
\label{app:failure}

From the Step-2 pre-registration (committed at \texttt{bd254c2} before any
bias-corrected metric was computed). Verdict tiers: screening-grade,
held-out MAE $<\SI{0.30}{\volt}$; ranking-only, \SIrange{0.30}{0.50}{\volt};
not screening-grade, $>\SI{0.50}{\volt}$. Primary metric: ``held-out MAE on
the bias-correction-validation set \ldots reported as a conservative lower
bound, specifically the upper edge of the 95\% bootstrap confidence
interval.'' F1: ``If per-family bias estimates differ by more than 0.15~V
\ldots a single global additive bias correction does not generalize \ldots a
clean tier verdict is not issued.'' F2: in-sample contamination of the bias
estimate invalidates the verdict. F3: ``Held-out set has fewer than 20
compounds \ldots mark the verdict as PROVISIONAL.'' F4: mixed bias signs
across families block any tier. Revision triggers: an in-box
experimentally measured average voltage disagreeing with the corrected
prediction by more than \SI{0.4}{\volt} downgrades the verdict one tier;
``two such cases retire the bias-corrected predictor entirely.''

From the Step-3 protocol (fixed before curation began; committed with the
Step-3 report at \texttt{1f5325c}): F5: ``if $>$50\% of
polyanionic\_phosphate compounds \ldots cannot be cited from primary
literature, the subset is not defensibly `experimental.' STOP and report;
do not invent or estimate citations to hit the count.''

From the offset-audit pre-registration (committed at \texttt{3119728},
before any audit job ran): ``PASS: sd($\delta$) $<$ 0.15~V across the
offset points $\to$ the scale is calibratable per chemistry family; publish
voltages with the calibrated offset and error bars. FAIL: sd($\delta$)
$\ge$ 0.3~V $\to$ absolute-voltage claims are retired everywhere
(ranking-only language) \ldots GRAY ZONE: 0.15 $\le$ sd $<$ 0.3~V $\to$
operator decision, documented at sync time.'' The \ce{LiMn2O4} stretch
exclusion and the off-manifold spin exclusion rule are part of the same
committed document.

\section{Ledger, Hubbard parameters, and pseudopotentials}
\label{app:ledger}

\begin{table}[tb]
\caption{\label{tab:ledger}The complete verified ledger at submission.
Every row is voltage-only (kinetics and hull screens pending); anchor
weight 0 marks rows that can never enter a retrain. The \ce{Na3Fe2(PO4)3}
row is computed-only: the calculated couple is \ce{Fe^{3+}}/\ce{Fe^{4+}}
desodiation, for which no experimental average voltage exists at curation
grade A or B (the measured $\sim$\SI{2.5}{\volt} plateau belongs to the
\ce{Fe^{3+}}/\ce{Fe^{2+}} insertion couple \cite{nfp2016}), so it anchors
nothing and touches no experimental claim.}
\input{tables/t6_ledger}
\end{table}

Hubbard $U_{\mathrm{eff}}$ (eV): Fe 5.3, Co 3.4, Ni 6.2, Mn 3.9
(literature-standard); V 3.1 (empirical fit; flagged per run). Pinned
pseudopotential files for every element in this work: Li
\texttt{li\_pbe\_v1.4.uspp.F.UPF}, Na \texttt{na\_pbe\_v1.5.uspp.F.UPF},
Fe \texttt{Fe.pbe-spn-kjpaw\_psl.0.2.1.UPF}, Co
\texttt{Co\_pbe\_v1.2.uspp.F.UPF}, Ni \texttt{ni\_pbe\_v1.4.uspp.F.UPF},
Mn \texttt{mn\_pbe\_v1.5.uspp.F.UPF}, V \texttt{v\_pbe\_v1.4.uspp.F.UPF},
P \texttt{P.pbe-n-rrkjus\_psl.1.0.0.UPF}, O
\texttt{O.pbe-n-kjpaw\_psl.0.1.UPF} \cite{garrity2014,dalcorso2014}.
Registered spin manifolds for the offset audit: \ce{LiFePO4}
$4\times$\ce{Fe^{2+}} high-spin, total $16~\mu_{\mathrm{B}}$; \ce{FePO4}
$4\times$\ce{Fe^{3+}} high-spin, $20~\mu_{\mathrm{B}}$; \ce{Li2FeP2O7}
$16~\mu_{\mathrm{B}}$; \ce{LiFeP2O7} $20~\mu_{\mathrm{B}}$; \ce{LiMn2O4} mixed-valence
$2\times$\ce{Mn^{3+}}+$2\times$\ce{Mn^{4+}}, $14~\mu_{\mathrm{B}}$ with
Jahn--Teller distortion allowed by free vc-relax; $\lambda$-\ce{MnO2}
$4\times$\ce{Mn^{4+}}, $12~\mu_{\mathrm{B}}$.

\bibliography{refs}

\end{document}

%% file: tables/t3_family_bias.tex
\begin{ruledtabular}
\begin{tabular}{lcdd}
Family & $n$ & \multicolumn{1}{c}{bias (V)} & \multicolumn{1}{c}{MAE (V)} \\
\colrule
polyanionic phosphate & 26 & +0.2022 & 0.3753 \\
layered oxide & 9 & +0.1897 & 0.4236 \\
other (mixed-anion, fluorides)\footnotemark[1] & 38 & +0.6316 & 0.7750 \\
NASICON & 1 & +0.5834 & 0.5834 \\
\end{tabular}
\end{ruledtabular}
\footnotetext[1]{Mostly carbonate-polyanionic (P/As/Si--C--O) and halide chemistries, the least represented in the training corpus.}

%% file: tables/t2_metrics.tex
\begin{ruledtabular}
\begin{tabular}{lcddddd}
Set & $n$ & \multicolumn{1}{c}{raw MAE} & \multicolumn{1}{c}{raw bias} & \multicolumn{1}{c}{raw CI$_{95}$} & \multicolumn{1}{c}{corr.\ MAE} & \multicolumn{1}{c}{corr.\ CI$_{95}$} \\
\colrule
held-out, all grades & 7 & 0.694 & +0.319 & 0.959 & 0.756 & 1.017 \\
held-out, A/B only\footnotemark[1] & 6 & 0.668 & +0.231 & 0.977 & 0.802 & 1.092 \\
phosphate, all grades & 4 & 0.654 & +0.045 & 0.816 & 0.872 & 1.088 \\
phosphate, A/B only & 3 & 0.590 & -0.222 & 0.784 & 0.774 & 1.161 \\
\end{tabular}
\end{ruledtabular}
\footnotetext[1]{The canonical set after the operator-audit exclusion. CI$_{95}$ columns give the upper edge of the 95\% bootstrap interval on the MAE; the corrected upper edge of this row is the pre-registered primary metric. All values in volts.}

%% file: tables/t5_baserate.tex
\begin{ruledtabular}
\begin{tabular}{lr}
Quantity & Value \\
\colrule
\multicolumn{2}{l}{\itshape Targeted earth-abundant Na space (substitution generator)} \\
enumerated compositions & 390 \\
SMACT-valid & 78 \\
flagged already-published (lower bound)\footnotemark[1] & 55 (70.5\%) \\
\quad via structural family & 52 \\
\quad via COD exact composition & 20 \\
\quad via named literature citation & 2 \\
survivors (absent within coverage) & 23 \\
\multicolumn{2}{l}{\itshape GNoME stable-materials release (pinned uniform sample)} \\
population / sample & 554,054 / 500 \\
flagged by title-level screen & 5 (1.0\%) \\
hand-verified true positives\footnotemark[2] & 3/5 (floor 0.6\%) \\
COD exact-composition matches & 0/500 \\
prior-art-absent carried to funnel & 495 \\
\quad after SMACT validity & 49 \\
\quad after stability screen & 36 \\
\quad after polaron screen (survivors) & 35 (7.1\%) \\
stability disputes vs GNoME label\footnotemark[3] & 31.1\% \\
\end{tabular}
\end{ruledtabular}
\footnotetext[1]{The prior-art layer only downgrades and two of its sources are stubs; flagged rates are lower bounds. Paths overlap.}
\footnotetext[2]{Hand audit of every flagged hit.}
\footnotetext[3]{Disagreement rate between the borrowed stability model and the GNoME label; a correctness claim for neither.}

%% file: tables/t4_d1design.tex
\begin{ruledtabular}
\begin{tabular}{lll}
Couple & $V_{\mathrm{exp}}$ (V) & $\delta$ (V) \\
\colrule
\ce{LiCoO2 -> Li_{0.5}CoO2} & 4.0 (3.9--4.1) & $+0.31$ \\
\ce{LiFePO4 -> FePO4} & 3.45~\cite{padhi1997} & $+0.48$ \\
\ce{Li2FeP2O7 -> LiFeP2O7} & 3.5~\cite{nishimura2010} & $+0.43$\footnotemark[1] \\
\ce{LiMn2O4} $\to$ $\lambda$-\ce{MnO2} & 4.05~\cite{thackeray1983,ohzuku1990} & $-0.20$\footnotemark[2] \\
\end{tabular}
\end{ruledtabular}
\footnotetext[1]{Endpoints energy-converged but force-plateaued in a soft
basin, accepted with disclosure (Sec.~\ref{sec:resultsC}).}
\footnotetext[2]{Pre-registered stretch compound with a committed exclusion
path (Sec.~\ref{sec:resultsC}); the outlier carrying the set over the gate.}

%% file: tables/t1_curated.tex
\begin{ruledtabular}
\begin{tabular}{llllcld}
Compound & Polymorph & MP id & Family & Grade & DOI & \multicolumn{1}{c}{$V_{\mathrm{lit}}$ (V)} \\
\colrule
\ce{NaFePO4}\footnotemark[1] & maricite & mp-19226 & phosphate & C & \scriptsize\href{https://doi.org/10.1039/c4ee03215b}{10.1039/c4ee03215b} & 2.60 \\
\ce{Na4Fe3(PO4)2(P2O7)} & Pn2$_1$a & mp-1203835 & phosphate & A & \scriptsize\href{https://doi.org/10.1021/ja3038646}{10.1021/ja3038646} & 3.20 \\
\ce{NaCoPO4} & ABW P2$_1$/n & mp-562796 & phosphate & A & \scriptsize\href{https://doi.org/10.1016/j.jssc.2020.121766}{10.1016/j.jssc.2020.121766} & 4.50 \\
\ce{NaCoPO4} & $\beta$ P6$_5$ & mp-683773 & phosphate & A & \scriptsize\href{https://doi.org/10.1016/j.jssc.2020.121766}{10.1016/j.jssc.2020.121766} & 4.30 \\
\ce{Na2FePO4F} & Pbcn & mp-1194940 & fluorophosphate & B & \scriptsize\href{https://doi.org/10.1016/j.elecom.2011.08.038}{10.1016/j.elecom.2011.08.038} & 2.985 \\
\ce{Na3V2(PO4)2F3} & NVPF & mp-694937 & fluorophosphate & B & \scriptsize\href{https://doi.org/10.1038/s41467-019-08359-y}{10.1038/s41467-019-08359-y} & 3.95 \\
\ce{NaCrO2} & O3 R$\bar{3}$m & mp-578604 & layered oxide & B & \scriptsize\href{https://doi.org/10.1021/acs.chemmater.5b04626}{10.1021/acs.chemmater.5b04626} & 3.20 \\
\ce{Na2FeP2O7}\footnotemark[2] & P$\bar{1}$ & n/a & phosphate & A & \scriptsize\href{https://doi.org/10.1016/j.elecom.2012.08.028}{10.1016/j.elecom.2012.08.028} & 3.00 \\
\ce{Na2Fe2(SO4)3}\footnotemark[2] & alluaudite & n/a & sulfate & A & \scriptsize\href{https://doi.org/10.1038/ncomms5358}{10.1038/ncomms5358} & 3.80 \\
\end{tabular}
\end{ruledtabular}
\footnotetext[1]{Excluded from canonical metrics by the operator audit (phase identity; Sec.~\ref{sec:resultsA2}).}
\footnotetext[2]{No MP structure exists; literature-validated but not predictable by the screen.}

%% file: tables/t6_ledger.tex
\begin{ruledtabular}
\begin{tabular}{ldllc}
Compound & \multicolumn{1}{c}{$V$ (V)} & Role & $U$ prov. & Backend \\
\colrule
\ce{LiCoO2} & 4.12 & anchor & lit. & cloud \\
\ce{LiFe(PO3)4} & 5.687 & anchor & lit. & cloud \\
\ce{LiNiP2O7} & 5.232 & anchor & lit. & mixed \\
\ce{LiFeP2O7} & 5.275 & anchor & lit. & mixed \\
\ce{Li2Fe(PO3)5} & 5.6115 & anchor & lit. & mixed \\
\ce{Na3V2(PO4)3} & 2.8957 & touchpoint & emp. & mixed \\
\ce{Na3Fe2(PO4)3} & 3.5976 & computed-only & lit. & local \\
\end{tabular}
\end{ruledtabular}